\documentstyle[]{article}
\advance\voffset by -2cm

\def\a{\alpha}
\def\b{\beta}
\def\c{\chi}
\def\d{\delta}
\def\e{\epsilon}

\def\f{\phi}

\def\j{\psi}
\def\k{\kappa}
\def\l{\lambda}
\def\m{\mu}

\def\q{\theta}

\def\D{\Delta}
\def\F{\Phi}
\def\G{\Gamma}

\def\S{\Sigma}

\def\co{{\cal O}}

\def\inbar{\vrule height1.5ex width.4pt depth0pt}
\def\rlx{\relax\leavevmode}
\def\I{\leavevmode\hbox{\small1\kern-3.8pt\normalsize1}}
\def\openone{\leavevmode\hbox{\small1\kern-3.3pt\normalsize1}}
\def\Ione{\rlx{\rm 1\kern-2.7pt l}}
\def\Ik{\rlx{\rm I\kern-.18em k}}
\def\IC{\rlx\leavevmode
             \ifmmode\mathchoice
                    {\hbox{\kern.33em\inbar\kern-.3em{\rm C}}}
                    {\hbox{\kern.33em\inbar\kern-.3em{\rm C}}}
                    {\hbox{\kern.28em\sinbar\kern-.25em{\rm C}}}
                    {\hbox{\kern.25em\ssinbar\kern-.22em{\rm C}}}
             \else{\hbox{\kern.3em\inbar\kern-.3em{\rm C}}}\fi}
\def\IP{\rlx{\rm I\kern-.18em P}}
\def\IR{\rlx{\rm I\kern-.18em R}}
\def\IN{\rlx{\rm I\kern-.20em N}}

\def\llsymbol#1{\@llsymbol{\@nameuse{c@#1}}}
\def\@llsymbol#1{\ifcase#1\or {}\or {'}\or {''}\or {'''}\or
   {''''}\or {'''''}\or  \else\@ctrerr\fi\relax}
\newcounter{contador}

\newcommand{\ol}\overline
\newcommand{\ti}\tilde
\newcommand{\wt}\widetilde
\newcommand{\wh}\widehat
\newcommand{\bv}\breve
\newcommand{\dg}\dagger

\newcommand{\aand}{\;\;\;\mbox{and}\;\;\;}

\newcommand{\be}{\begin{equation}}
\newcommand{\ee}{\end{equation}}
\newcommand{\bl}{\begin{eqnarray}&}
\newcommand{\el}{&\end{eqnarray}}
\newcommand{\bq}{\begin{eqnarray}}
\newcommand{\eq}{\end{eqnarray}}

\newcommand{\pa}{\partial}

\begin{document}
{\hfill\parbox{45mm}{{\large hep-th/9712100\\
                             CBPF-NF-053/97\\
                             UFES-DF-OP97/4}} \vspace{3mm}

\begin{center}
{{\Large {\bf Renormalization of the $N=1$ Abelian
       \\[2mm]Super-Chern-Simons Theory Coupled to\\[4mm] 
              Parity-Preserving Matter}}} 
\vspace{7mm} 


{\large L.P. Colatto$^{{\rm (a),}}$\footnote{Current address: {\it Centro 
Brasileiro de Pesquisas F\'\i sicas (CBPF), Departamento de
Teoria de Campos e Part\'\i culas (DCP), Rua Dr. Xavier Sigaud 150 - 
22290-180 - Rio de Janeiro - RJ - Brazil}.}, 
M.A. De Andrade$^{{\rm (b),(c)}}$, O.M. 
Del Cima$^{{\rm (b),(c),}}$\footnote{Current address: 
{\it Institut f\"ur Theoretische Physik, 
Technische Universit\"at Wien, Wiedner Hauptstra{\ss}e 8-10 - 
A-1040 - Vienna - Austria}.},\\[2mm] D.H.T. Franco$^{{\rm 
(c),}}$\footnote{Supported by the {\it Conselho Nacional de Desenvolvimento 
Cient\'\i fico e Tecnol\'ogico (CNPq)}.}, J.A. Helay\"el-Neto$^{{\rm (c)}}$ 
and O. Piguet$^{{\rm (d),3,}}$\footnote{Supported in part by the {\it Swiss 
National Science Foundation}.}${}^{,}$\footnote{On leave of 
absence from {\it D\'epartement de Physique Th\'eorique, 
Universit\'e de Gen\`eve, 24 quai E. Ansermet - CH-1211 - Gen\`eve 4 - 
Switzerland}.} } 
\vspace{4mm}

$^{{\rm (a)}}$ {\it International Centre for Theoretical Physics (ICTP), 
\\High Energy Section (HE), \\Strada Costiera 11 - P.O. Box 586 - 34100 - 
Trieste - Italy.} 

$^{{\rm (b)}}$ {\it Pontif\'\i cia Universidade Cat\'olica do Rio de Janeiro
(PUC-RIO), \\Departamento de F\'\i sica, \\Rua Marqu\^es de S\~ao Vicente, 
225 - 22453-900 - Rio de Janeiro - RJ - Brazil.}

$^{{\rm (c)}}$ {\it Centro Brasileiro de Pesquisas F\'\i sicas (CBPF), 
\\Departamento de
Teoria de Campos e Part\'\i culas (DCP),\\Rua Dr. Xavier Sigaud 150 - 
22290-180 - Rio de Janeiro - RJ - Brazil.}

$^{{\rm (d)}}$ {\it Universidade Federal do Esp\'\i rito Santo (UFES), 
\\
Departamento de F\'\i sica, \\Campus Universit\'ario de Goiabeiras - 
29060-900 - Vit\'oria - ES - Brazil.}
\vspace{4mm} 

{\it E-mails: colatto@cbpfsu1.cat.cbpf.br, marco@cbpfsu1.cat.cbpf.br, 
delcima@tph73.tuwien.ac.at, dfranco@cbpfsu1.cat.cbpf.br, 
helayel@cbpfsu1.cat.cbpf.br, piguet@cce.ufes.br.} 
\end{center}

\begin{abstract}
We analyse the renormalizability of an Abelian $N$$=$$1$ super-Chern-Simons
model coupled to parity-preserving matter on the light of the 
regularization
independent algebraic method. The model shows to be stable under 
radiative
corrections and to be gauge anomaly free.
\end{abstract}

\newpage
\section{Introduction}
Our purpose in this letter is to investigate the renormalizability of the 
$N$$=$$1$ super-Chern-Simons theory~\cite{super-cs}  
coupled in a parity-preserving way with matter 
supermultiplets~\cite{andcima}, by using the method of 
algebraic renormalization~\cite{pigsor}. The latter is based on the 
BRS-formalism~\cite{brs} which together with the Quantum Action 
Principle~\cite{qap} lead to a regularization independent scheme. 
The model considered here comes 
from an 
$N$$=$$1$ super-QED in the Atiyah-Ward space-time~\cite{andcima} after a 
dimensional 
reduction proposed by Nishino in ref.~\cite{nishino} and suitable 
truncations of some spurious degrees of freedom caused by dimensional
reduction of a time coordinate.  

\section{The model and its symmetries}
The gauge invariant action for the $N$$=$$1$ super-Chern-Simons 
theory  
coupled in a parity-preserving way with matter supermultiplets, 
in superspace\footnote{The notations and
 conventions adopted throughout the work are those of ref.~\cite{gates}.}, is 
given by~\cite{super-cs,andcima}: 
\bq
\Sigma _{inv}\!\!\!&=&\!\!\! \frac{1}{2} \int dv\left\{ \k(\G^\a W_\a)-(
\nabla^\a \bar{\F} _{+})(\nabla_\a \F_{+})-(\nabla^\a \bar{\F}
_{-})(\nabla_\a \F_{-})+m(\bar{\F} _{+}\F_{+}-\bar{\F} _{-}\F_{-})\,+\right.  
\nonumber \\  
&-&\left.\!\!\! \l (\bar{\F}_{+}\F_{+}-\bar{\F}_{-}\F_{-})^2 \right\}\,.  
\label{inv} 
\eq
The coupling of matter 
coupling here is the same as the one occuring  in 
the super-$\tau_3$ QED model of ref.~\cite{andcima}.

The Chern-Simons parameter $\k$ is the inverse of the coupling 
constant in term of which perturbation expansion is defined. The 
superspace measure adopted is $dv$ $\equiv$ $d^3xd^2\q$. 
The gauge superconnection is a spinorial Majorana superfield $\G_\a$, and
matter is represented by the  complex scalar superfields  $\F_{\pm}$,
with opposite $U(1)$-charges. 
The covariant spinorial derivatives are defined as follows: 
\be
\nabla_\a\F_{\pm}=\left( D_\a\mp i\G_\a\right) \F_{\pm}\;\;\;
\mbox{and}\;\;\;\nabla_\a \bar{\F}_{\pm}=\left({D}_\a\pm i{
\G}_\a\right) \bar{\F}_{\pm}\;\;\;\;,  \label{covder}
\ee
where $D_\a=\pa_\a+i \q^\b \pa_{\a\b}$. The superfield-strength $W_\a$ is 
given by
\be
W_\a=\frac{1}{2} D^\b D_\a \G_\b \;.
\label{strength}
\ee
The component-field projections read 
\bq
&&  \c_\a = \G_\a | ~~,~~~B' = \frac{1}{2} D^\a \G_\a | ~~,\nonumber \\
&&  V_{\a\b}= -\frac{i}{2}D_{(\a}\G_{\b)} | ~~,~~~\l_\a = \frac{1}{2} D^\b 
D_\a \G_\b | ~~.  \label{gsc}
\eq
and
\bq
&&A_{\pm}(x)=\F_{\pm }(x,\q)|  \aand \bar{A}_{\pm}(x)=
\bar{\F}_{\pm }(x,\q)| \, , \nonumber \\
&& \j^\a_{\pm}(x) = D^\a \F_{\pm }(x,\q)| \aand  
\bar{\j}^\a_{\pm}(x) = D^\a \bar{\F}_{\pm }(x,\q)| \,,\nonumber \\
&&F_{\pm}(x) = D^2 \F_{\pm }(x,\q)| \aand  
\bar{F}_{\pm}(x) = D^2 \bar{\F}_{\pm }(x,\q)| \, .
\label{expan}
\eq

We choose a gauge-fixing action of the Landau type:
\bq
\Sigma _{gf}={\frac 12}~s\int{dv}~{\wh C}D^\a\G_\a=
\int{dv}\left\{{\frac 12}B D^\a\G_\a + {\wh C}D^2 C \right\}
 \;\;,
\label{gf}
\eq
with $C$, $\wh C$ and $B$ being respectively, the 
ghost, the antighost and the
Lagrange multiplier superfields.

The action, $\S_{inv}+\S_{gf}$, is invariant under the
following BRS transformations:
\bq
&&s\F_\pm=\pm iC\F_\pm ~~,~~~ s\bar{\F}_\pm=\mp 
iC\bar{\F}_\pm~~, \nonumber\\ 
&&s{\G}_\a={D}_\a C ~~,~~~ s{\wh C}= B ~~, \nonumber\\ 
&&sC=0 ~~, ~~~sB=0 ~~.   
\label{BRS}
\eq

In view of expressing the BRS-invariance of the model in a functional way
by a Slavnov-Taylor identity, we add to the action a source term, 
$\S_{ext}$, which contains external sources $J_\pm$, $\bar J_\pm$
coupled to the non-linear 
BRS variations of the quantum fields. The external coupling then reads 
\be
\S_{ext}=\int {dv}\left\{ \bar{J}_{+}\,s\F
_{+}+\bar{J}_{-} \,s\F_{-}
+s\bar{\F} _{+}\,J_{+}+s\bar{\F}
_{-}\,J_{-}\right\} ~~.  \label{ext}
\ee

Then, the Slavnov-Taylor identity obeyed by the complete action 
\be
\S=\S_{inv}+\S_{gf}+\S_{ext}~~,  \label{total}
\ee
is given by  
\bq
{\cal S}(\Sigma)\!\!\!&=&\!\!\!\int {dv}\left\{ 
D^\a C{\frac{\d 
\S}{\d \G^\a}}+B{\frac{\d \S}{\d {\wh C}}}+
\frac{\d \S}{\d \bar{J}_{+}}
{\frac{\d \Sigma}{\d
\F_{+}}}+\frac{\d \S }{\d \bar{J}_{-}}{\frac{\d
\S}{\d \F_{-}}}~+\right. \nonumber\\
&-&\left.\!\!\!\frac{\d \S}{\d J_{+}}\frac{\d
\S}{\d \bar{\F}_{+}}
-{\frac{\d \S}{\d J_{-}}}
\frac{\d \S}{\d \bar{\F}_{-}}\right\} =0~~.
\label{slavnov}
\eq

The corresponding linearized Slavnov-Taylor operator reads
\bq
{\cal S}_\S \!\!\!&=&\!\!\!\int {dv}\left\{D^\a C{\frac{\d
\S}{\d \G^\a}}+B{\frac{\d \S}{\d {\wh C}}}+
{\frac{\d \S}{\d \bar{J}_{+}}}{\frac \d {\delta \F_{+}}}+{\frac{\d \S}{\d 
\bar{J}_{-}}}{\frac \d {\d 
\F_{-}}}+{\frac{\d \S}{\d \F_{+}}}
{\frac \d {\d \bar{J}_{+}}}+{\frac{\d \S}{\d \F_{-}}}{\frac \d 
{\d \bar{J}_{-}}}\;+\right.  \nonumber \\ 
&-&\left.\!\!\!{\frac{\d \S}{\d J_{+}}}{\frac \d {\delta\bar{\F}_{+}}}-
{\frac{\d 
\S}{\d J_{-}}}{\frac \d {\d \bar{\F}_{-}}}-{\frac{\d \S}{\d 
\bar{\F}_{+}}}{\frac \d 
{\d J_{+}}}-{\frac{\d \S}{\d \bar{\F}_{-}}}{\frac 
\d {\d J_{-}}}\right\} ~~.  \label{slavnovlin}
\eq
The operation of ${\cal S}_{\S}$ over the fields and the external
sources is given by 
\bq
&&{\cal S}_{\S}\f=s\f
~~,~~~\f=\F_{\pm},\;\bar{\F}_{\pm},\;\G_\a,\;C,\;\wh C \!\!\!\aand\!\!\! 
B ~~,  \nonumber \\
&&{\cal S}_{\S}J_+=-{\frac{\d\S}{\d \bar{\F}_+}} ~~,~~~ 
{\cal S}_{\S}J_-=-{\frac{\d\S}{\d \bar{\F}_-}} ~~,  \nonumber\\ 
&&{\cal S}_{\S}\bar{J}_+={\frac{\d\S}{\d \F_+}} 
~~,~~~{\cal S}_{\S}\bar{J}_-={\frac{\d\S}{\d \F_-}}\;\;\;. 
\label{operation1}
\eq

The gauge condition, the ghost equation and the antighost equation 
~\cite{pigsor} for (\ref{total}) read 
\bq
&&{\frac{\d \S}{\d B}}={\frac 12} D^\a
\G_\a ~~,  \label{gaugecond} \\ 
&&{\frac{\d \S}{\d \wh C}}=D^2C ~~,
\label{ghostcond} \\ 
&&-i{\frac{\d \S}{\d C}}=\D_{{\rm class}
}~~,  \label{antighostcond} \\ 
&&\D_{{\rm class}}=iD^2\wh 
C-\bar{J}_{+}\,\F_{+}+\bar{J}_{-}\,\F_{-}-\bar{\F}_{+} \,J_{+} + 
\bar{\F}_{-}\,J_{-}~~. \nonumber
\eq
Note that the right-hand sides being linear in the 
quantum fields, will not be submitted to renormalization. 

Moreover, from the antighost equation (\ref{antighostcond}) and from the
Slavnov-Taylor identity (\ref{slavnov}) we get the $U(1)$ rigid invariance
\be
{\cal W}_{{\rm rigid}}\S =0~~,  \label{crigidcond}
\ee
where ${\cal W}_{{\rm rigid}}$ is the Ward operator of  $U(1)$ rigid symmetry
defined by 
\be
{\cal W}_{{\rm rigid}}=\int {dv}\left\{ \F_{+}
{\frac \d {\d \F_{+}}}-\F_{-}{\frac \d {\d \F_{-}}}-
\bar{\F}_{+}\frac \d {\d \bar{\F}_{+}}+\bar{\F}_{-} 
\frac \d {\d \bar{\F}_{-}}\right\} ~~. 
\label{wrigid}
\ee

The Slavnov-Taylor identity (\ref{slavnov}), the constraints 
(\ref{gaugecond}--\ref{antighostcond}) and the rigid invariance 
(\ref{crigidcond}) form an algebra that takes the form: 
\bq
&&{{\cal S}_{{\cal F}}{\cal S}({\cal F})=0}~~,~~~\forall~{\cal F}~~,  
\label{fcond1} \\
&&{{\cal S}_{{\cal F}}{\cal S}_{{\cal F}}=0}~~~{\mbox{if}}~~~{{\cal S}
({\cal F})=0}~~,  \label{fcond2} \\
&&{\frac{\d {\cal S}({\cal F})}{\d B}}-{\cal S}_{{\cal F}}\left({
\frac{\d {\cal F}}{\d B}}-\frac 12D^\a\G_\a\right) =
\left( {\frac{\d {\cal F}}{\d  \wh C}}-{D^2 C}\right)~~, 
\label{fcond3} \\
&&{\frac{\d {\cal S}({\cal F})}{\d \wh C}}+{\cal S}_{{\cal F}}
{\frac{\d {\cal F}}{\d \wh C}}=0~~,  \label{fcond4} \\ 
&&-i\int 
dv\frac \d {\d C}{\cal S}({\cal F})+{\cal S}_{{\cal F}}\int 
dv\left( -i\frac \d {\d C}{\cal F}-\D_{{\rm class}}\right) = 
{\cal W}_{{\rm rigid}}{\cal F}~~,  \label{fcond5} \\ 
&&{\cal W}_{{\rm 
rigid}}{\cal S}({\cal F})-{\cal S}_{{\cal F}}{\cal W}_{{\rm rigid}} {\cal 
F}=0~~,  \label{fcond6} 
\eq
where ${\cal F}$ is an arbitrary functional 
of ghost number zero. 

Finally, the action (\ref{total}) is invariant under the discrete 
symmetries displayed below:
\begin{enumerate}
\item[] Parity: 
\begin{equation}\begin{array}{llll}
\F_{\pm} \longleftrightarrow  \bar{\F}_{\pm}\ ,\quad
&J_{\pm} \longleftrightarrow  -\bar{J}_{\pm}\ , &&\\ 
\G_\a  \longrightarrow  -\G_\a\ ,\quad &B \longrightarrow -B\ ,\quad
&C  \longrightarrow -C\,\quad \wh C \longrightarrow -\wh C\ .
\end{array}\label{ds1}\end{equation}
\item[]$G$-parity:
\begin{equation}\begin{array}{llll}
\F_{\pm} \longrightarrow  -\F_{\pm}\ ,\quad 
&\bar{\F}_{\pm}\longrightarrow  -\bar{\F}_{\pm}\ ,\quad 
&J_{\pm} \longrightarrow  -J_{\pm}\ ,\quad
&\bar{J}_{\pm} \longrightarrow  -\bar{J}_{\pm}\ , \\ 
\G_\a  \longrightarrow  \G_\a\ ,\quad 
&B  \longrightarrow  B\ , \quad
&C  \longrightarrow  C \ ,\quad 
&\wh C \longrightarrow  \wh C \ .
\end{array}\label{ds2}\end{equation}
\end{enumerate}

The ultraviolet and infrared dimensions, $d$ and $r$ respectively, as well
as the ghost numbers, $\F\Pi$, and the Grassmann parity, $GP$, of all
superfields and superspace objects ($\q_\a$ and $D_\a$) are
collected in Table~\ref{table1}.
\begin{table}[hbt]
\centering
\begin{tabular}{|c||c|c|c|c|c|c|c|c|}
\hline
& $\G_\a$ & ${\F_\pm}$ & $C$ & ${\wh C}$ & $B$ & ${J_\pm}$ & $
\q_\a $ & $D_\a$ \\ \hline\hline
$d$ & ${1/2}$ & ${1/2}$ & 0 & 1 & ${1}$ & ${3/2}$ & $-{1/2}$ & ${1/2}$ \\ 
\hline
$r$ & ${1/2}$ & ${3/2}$ & 0 & 1 & ${1}$ & ${3/2}$ & $-{1/2}$ & ${1/2}$ \\ 
\hline
$\F\Pi$ & 0 & 0 & 1 & $-1$ & 0 & $-1$ & 0 & 0 \\ \hline
$GP$ & 1 & 0 & 1 & 1 & 1 & 1 & 1 & 1 \\ \hline
\end{tabular}
\caption[t1]{UV and IR dimensions, $d$ and $r$, ghost numbers, $\F\Pi$,
and Grassmann parity, $GP$.}
\label{table1}
\end{table}

\section{Renormalization}
The renormalizability of the model runs according to the usual procedure of
studying its stability -- which amounts to check that, 
in the quantum theory, the possible counterterms can be reabsorbed by a 
redefinition of the initial
parameters of the model -- and the determination of the possible anomalies.

\subsection{Stability}
In order to study the stability of the model under radiative corrections, we
introduce an infinitesimal perturbation in the classical action $\Sigma $ by
means of an integrated local functional $\S^c$ -- which, in the 
perturbative quantum theory,  would be considered as an infintesimal 
quantum correction, i.e. a counterterm --
\be
\S \rightarrow \S +\e \S^c ~~,  \label{stabe}
\ee
where $\e$ is an infinitesimal parameter. The functional $\S^c$
has the same quantum numbers as the action in the tree approximation.

The perturbed action must satisfy, to the order $\e$, the same
equations as $\S$, i.e., the Slavnov-Taylor identity, the gauge
condition, the ghost and antighost equations together with rigid invariance.
This implies the conditions
\be
{\cal S}_\S \S^c=0 ~~,  \label{stabcond}
\ee
\be
{\frac{\d\S^c}{{\d B}}}=0~~,~~~{\frac{\d \S^c}{
{\d {\wh C}}}}=0~~,~~~{\frac{\d \S^c}{{\d C}}}
=0~~,  \label{suplcond}
\ee
\be
{\cal W}_{{\rm rigid}} \S^c=0~~.  \label{crigidcond1}
\ee
Notice that $\S^c$ 
has also to be invariant under the discretes symmetries (\ref{ds1}--\ref
{ds2}).

The condition (\ref{stabcond}), due to nilpotency of the linearized 
Slavnov-Taylor operator, ${\cal S}_\S$, constitutes a cohomology 
problem in the sector of ghost number zero. The most general invariant 
counterterms, $\S^c$, solution of the constraints 
(\ref{stabcond}--\ref{crigidcond1}), are given by the invariant 
pieces of the 
classical action (\ref{inv}):\footnote{Note that the 
combination of counterterms given by 
\[
\frac12 {\cal S}_\Sigma \int dv\left( \bar J_+\Phi_+ +  \bar J_-\Phi_-
- \bar\Phi_+ J_+ - \bar\Phi_- J_- \right)
 = -\Sigma^c_2 + m\Sigma^c_3 - 2\l \Sigma^c_4\ ,
\]
is ``trivial'' , being a ${\cal S}_\Sigma$ - variation. The corresponding 
coefficient -- which may be fixed by the second normalization condition 
(\ref{norm-cond}) -- corresponds to the (nonphysical) renormalization of 
the matter field amplitude. The other three counterterms correspond to 
the renormalization of the physical parameters $\kappa$, $m$ and 
$\lambda$.}
\bq
&&\Sigma^c_1 = \int dv~\G^\a W_\a~,~~ 
\Sigma^c_2 = \int dv\left\{(\nabla^\a \bar{\F} _{+})
   (\nabla_\a \F_{+})+(\nabla^\a \bar{\F}_{-})
     (\nabla_\a \F_{-})\right\}~,~~\nonumber \\
&&\Sigma^c_3 = \int dv~(\bar{\F} _{+}\F_{+}-\bar{\F} _{-}\F_{-})~,~~
\Sigma^c_4 = \int dv~(\bar{\F}_{+}\F_{+}-\bar{\F}_{-}\F_{-})^2~. \label{cterm}
\eq
The corresponding coefficients are fixed by suitable normalization 
conditions, which may be chosen as below:  
\begin{equation}\begin{array}{ll}
\left. C_{\b\a}D^2(p){\frac \pa {\pa p^2}}\G_{\G\G}^{\a\b}\right|^
{\q=0}_{p=p(\m)} = {\frac 12}~k~~, \qquad
&{\frac 12}\left. D^2(p){\frac \pa {\pa p^2}}
\G_{\bar{\F}_{\pm}\F_{\pm}}\right|^{\q=0}_{p=p(\m)} = -
{\frac 12}~~,   \\[3mm]
\left. \G_{\bar{\F} _{\pm}\F_{\pm}}\right|^{\q=0}_{p^2=m^2}
 = 0~~,  \qquad
&\left. \G_{\bar{\F}_{\pm}\F_{\pm}\bar{\F}_{\pm}\F_{\pm}}\right|^
{\q=0}_{p=p(\m)}  = -\lambda~~, 
\end{array}\label{norm-cond}\end{equation}
where $\m$ is an energy scale, $p(\m)$ some reference set of
4-momenta at this scale and $D_\a(p)=\pa_\a+\q^\b p_{\a\b}$ .

\subsection{Anomaly}
Here, our goal, in order to complete the proof of the renormalizability of 
the model, is to show that it is possible to extend the symmetry 
preserving perturbative expansion to all orders, i.e., we must prove that 
is possible to define a quantum vertex functional $\G$: 
\be
\G=\S +\co(\hbar) ~~,  \label{funvert}
\ee
such that
\begin{equation}
{{\cal S}(\G)=0}~~,  
\label{qslavnov}\end{equation}
and
\begin{equation}
{\frac{\d\G}{\d B}}={\frac 12} D^\a\G_a~~, \quad
{\frac{\d \G}{\d \wh C}}=D^2C~~, \quad
-i{\frac{\d \G}{\d C}}=\D_{{\rm class}}~~,\quad 
{\cal W}_{{\rm rigid}}\G=0~~. 
\label{qgaugecond}\end{equation}

According to the Quantum Action Principle~\cite{qap}, the Slavnov-Taylor
identity (\ref{slavnov}) will be broken at the quantum level, as follows
\be
{\cal S}(\G)=\D \cdot\G =\D 
+\co(\hbar\D)~~.  \label{breaking}
\ee
where $\D$, at lowest order in $\hbar$ ,is an integrated local
functional with ghost number 1, UV dimension $\leq {2}$ and IR dimension 
$\geq {2}$.

By using the algebra (\ref{fcond1}--\ref{fcond6}) written for the functional 
$\G$ we get the following set of constraints for the breaking $\D$:  
\begin{equation}
{{\cal S}_\S \D}=0~~,  
\label{breakcond1}\end{equation}
and
\begin{equation}
\frac{\d \D}{\d B}=0~~,  \quad
\frac{\d \D}{\d \wh C}=0~~,  \quad
\int dv\frac \d {\d C}\D=0~~,   \quad
{\cal W}_{{\rm rigid}}\D=0~~. 
\label{breakcond2}\end{equation}

The condition (\ref{breakcond1}) represents the Wess-Zumino consistency
condition. It
constitutes a cohomology problem analogous to the one determining the
counterterm $\S^c$, now in the sector of ghost number one. Its solution
can always be written as a sum of a trivial cocycle 
${\cal S}_\S {\wh\D}^{(0)}$, where ${\wh\D}^{(0)}$ has ghost number $0$, 
and of
nontrivial elements belonging to the cohomology of ${\cal S}_\S$ (\ref
{slavnovlin}) in the sector of ghost number one: 
\be
\D^{(1)}={\cal A}^{(1)}+{\cal S}_\S {\wh\D}^{(0)}~~.
\label{breaksplit}
\ee
The trivial cocycle ${\cal S}_\S {\wh\D}^{(0)}$ can be absorbed
into the vertex functional $\G$ as a noninvariant integrated local
counterterm $-{\wh\D}^{(0)}$. On the other hand, a nonzero 
${\cal A}^{(1)}$ would represent a possible anomaly. 

Considering the third of the conditions (\ref{breakcond2}), 
which $\D^{(1)}$  has to satisfy, it can be concluded that the latter 
 has the form
\be
\D^{(1)}=\int {dv}~K^{\a (0)} D_\a C ~~.
\label{anomaly}
\ee
By analyzing the Slavnov-Taylor operator ${\cal S}_\S$ (\ref{slavnovlin}) 
and the Wess-Zumino consistency condition (\ref{breakcond1}), one sees
that the breaking $\D^{(1)}$ has UV and IR dimensions bounded by $d
\leq {2}$ and $r \geq 2$. Therefore, the dimensions of $K^{\a (0)}$ must be
bounded by $d \leq \frac 32$ and $r \geq \frac 32$ .
It has ghost number $0$.

Now, after solving all the conditions $K^{\a (0)}$ has to fulfill, we  
see that it may be expanded in a local basis as:                
\be
K^{\a (0)}=\sum_{i=1}^3 a_i~K_i^{\a (0)}~~,
\label{kas}
\ee
where
\be
K_1^{\a (0)}=\G^\a\G^2~~,~~~K_2^{\a (0)}=\bar\F_{+}\G^\a\F_{+}~~,~~~
K_3^{\a (0)}=\bar\F_{-}\G^\a\F_{-}~~.  \label{kas1}
\ee
We may thus write
\be
K^{\a (0)}D_\a C=\sum_{i=1}^3 a_i~K_i^{\a (0)}D_\a C=
\sum_{i=1}^3 a_i~{\cal S}_\S{\wh K}_i^{(0)}={\cal S}_\S{\wh K}^{(0)}~~,
\label{tkas}
\ee
where
\be
{\wh K}_1^{(0)}=-{\frac 12}(\G^2)^2~~,~~~{\wh K}_2^{(0)}=-\bar\F_{+}\G^2 
\F_{+}~~,
~~~{\wh K}_1^{(0)}=-\bar\F_{-}\G^2 \F_{-}~~.  \label{kas11} 
\ee
Therefore, the breaking $\D^{(1)}$ is BRS trivial and given by
\be
\D^{(1)}=\int {dv}~K^{\a (0)} D_\a C ={\cal S}_\S\int {dv}~{\wh K}^{(0)}
={\cal S}_\S {\wh\D}^{(0)}~~.  
\label{kas2}
\ee
This means that ${\cal A}^{(1)}=0$ in 
(\ref{breaksplit}),
which implies the implementability of the Slavnov-Taylor identity to every
order through the absorbtion of the noninvariant counterterm 
$-{\wh\D}^{(0)}$ .

Of course, invariant counterterms may still be arbitrarily added at each
order. However the result of the discussion on the stability of the
classical theory shows that these counterterms correspond to a
renormalization of the parameters of the theory. Their coefficients 
are fixed by the normalization conditions (\ref{norm-cond}).

In conclusion, the use of the algebraic method of renormalization has 
allowed us to show
that $N$$=$$1$ super-Chern-Simons model coupled to parity-preserving matter
is perturbatively renormalizable to all orders. First, the study of the
stability has led the conclusion that counterterms can be reabsorbed by a
redefinition of the initial parameters of the model. Next, gauge anomalies
have been proven to be absent. We stress that this is a 
purely algebraic analysis, valid to all orders and does not involve any 
regularization scheme, nor any Feynman graph calculation. 

\subsection*{Acknowledgements} 
L.P.C., O.M.D.C. and D.H.T.F. thank to the 
{\it High Energy Section} of the {\it ICTP - Trieste - Italy}, where 
part of this work was done, for the kind 
hospitality and financial support, and to its Head, Prof. S. 
Randjbar-Daemi. CNPq-Brazil is acknowledged for invaluable financial help. 
One of the authors (O.M.D.C.) dedicates 
this work to his wife,
Zilda Cristina, to his daughter, Vittoria, and to the other baby 
who is coming.

\end{document}